# Human vs. LLM-Based Thematic Analysis for Digital Mental Health Research: Proof-of-Concept Comparative Study (2025)

Karisa Parkington*, Bazen G. Teferra*, Marianne Rouleau-Tang, Argyrios Perivolaris, Alice Rueda, *IEEE Member,* Adam Dubrowski, Bill Kapralos, Reza Samavi, Andrew Greenshaw, Yanbo Zhang, Bo Cao, Yuqi Wu, Sirisha Rambhatla, Sridhar Krishnan, *IEEE Member,* & Venkat Bhat

*Abstract*—Thematic analysis offers rich insights into participants' experiences through coding and theme development, but its resource-intensive nature limits application in larger healthcare studies. Large language models (LLMs) can analyze text at scale and identify key content automatically, potentially overcoming these barriers. However, their application to mental health interviews requires direct comparison with traditional human-based analysis. This proof-of-concept study evaluates out-of-the-box and knowledge-base LLM-based thematic analysis against traditional human processes using semi-structured interview transcripts from a stress-reduction trial with healthcare workers. OpenAI's GPT-4o model was used based on the Role, Instructions, Steps, End-Goal, Narrowing (RISEN) prompt engineering framework and compared to human analysis conducted in Dedoose. Each approach independently developed codes, noted saturation points, applied codes to excerpts for a subset of participants (n = 20), and synthesized data into themes. Outputs and performance metrics were directly compared. GPT-4o LLMs executing the RISEN framework for qualitative thematic synthesis developed deductive parent codes comparable to human-derived parent codes. However, humans provided superior detail in inductive child code development and theme synthesis. Knowledge-based LLMs achieved coding saturation with fewer transcripts (10-15) compared to the out-of-the-box model (15-20) and human (90-99). The out-of-the-box LLM identified a comparable set of excerpts to human researchers with strong inter-rater reliability (K = 0.84), although the knowledge-base LLM produced markedly fewer excerpts. Human excerpts tended to be longer, with multiple codes per excerpt, than the LLMs, which typically applied one code per excerpt and could not replicate reliability measures. Overall, LLM-based thematic analysis was more cost-effective but lacked the specificity and depth of human analysis. LLMs show the potential to transform qualitative analysis in mental healthcare and clinical research when combined with human oversight (human-AI collaboration), enabling consideration of participants' perspectives while balancing research resources.

*Index Terms*— GPT, digital health, human comparative study, large language models (LLMs), psychiatry, semi-structured interviews, thematic analysis

## I. INTRODUCTION

P ATIENT perspectives are fundamental to optimizing patient-centred care, especially in pilot, feasibility, and implementation-effectiveness trials [1-3]. In the psychiatry and mental health, virtual reality (VR) and digital interventions are increasingly used, yet their feasibility and effectiveness are often assessed through quantitative methods. Unlike quantitative methods, qualitative research captures subjective thoughts, emotions, and behaviours, offering nuanced data essential for healthcare decision-making [4-6]. For instance, reflexive thematic analysis is a widely-used methodology providing rich qualitative insights into patient experiences and user perspectives through the systematic

This work was supported in part by the Canadian Department of Defense under Grant CovCA-0592 and by a Postdoctoral Research Fellowship awarded to K. Parkington and B.G. Teferra by the Canadian Institute for Health Research. Co-first authors: K. Parkington and B.G.Teferra. Corresponding author: V. Bhat.

K. Parkington (email: karisa.parkington@unityhealth.to), B. G. Teferra (email: bazengashaw.teferra@unityhealth.to), M. Rouleau-Tang (email: Marianne.rouleau-tang2@unityhealth.to), A. Perivolaris (email: argyrios.perivolaris@unityhealth.to), and A. Rueda (email: alice.rueda@unityhealth.to) are affiliated with the Interventional Psychiatry Program at St. Michael's Hospital, Unity Health Toronto (Toronto, Ontario, Canada, M5B 1M4).

A. Dubrowski (email: adam.dubrowski@ontariotechu.net) and B. Kapralos (email: bill.kapralos@ontariotechu.net) are affiliated with the maxSIMhealth Group at Ontario Tech University (Oshawa, Ontario, Canada, L1G 0C5).

R. Samavi (email: samavi@torontomu.ca) and S. Krishnan (email: krishnan@torontomu.ca) are affiliated with the Department of Electrical, Computer, and Biomedical Engineering at Toronto Metropolitan University (Toronto, Ontario, Canada, M5B 2K3).

A. Greenshaw (email: agreensh@ualberta.ca), Y. Zhang (email: yanbo9@ualberta.ca), and B. Cao (email: cloud.cao@ualberta.ca) are affiliated with the Faculty of Medicine and Dentistry at the University of Alberta (Edmonton, Alberta, Canada, T6G 2R7).

Y. Wu (email: yuqi14@ualberta.ca) is affiliated with the Department of Computer Engineering, Faculty of Engineering at the University of Alberta (Edmonton, Alberta, Canada, T6G 1H9).

S. Rambhatla (email: sirisha.rambhatla@uwaterloo.ca) is affiliated with the Department of Management Science and Engineering, Faculty of Engineering at the University of Waterloo (Waterloo, Ontario, Canada, N2L 3G1).

V. Bhat (email: venkat.bhat@utoronto.ca) is affiliated with the Department of Psychiatry, Temerty Faculty of Medicine at the University of Toronto (Toronto, Ontario, Canada, M5S 1A8) and the Interventional Psychiatry Program at St. Michael's Hospital, Unity Health Toronto (Toronto, Ontario, Canada, M5B 1M4).

Supplementary 1 presents the RISEN prompts used for each step of the LLM thematic analysis process. Supplementary 2 presents code breakdowns and comparisons across human and LLM methods.

Colour versions of one or more of the figures in this article are available online at http://ieeexplore.ieee.org

identification, analysis, and reporting of themes within text-based datasets, such as semi-structured interview transcripts [7]. Braun and Clarke [7] overview the four key steps to thematic analysis in psychology: 1) iterative parent (superordinate) and child (nested) code development using deductive and/or inductive approaches until saturation is reached, 2) operationally defining codes in a manual to ensure consistency, 3) identifying, extracting, and systematically application of codes (labels or tags assigned to specific text excerpts) to excerpts in text, and 4) between-participant theme synthesis to identify overarching patterns in the data. By defining and applying codes (deductive or inductive labels or tags) to categorize and organize text data into recurring patterns relevant to the research question(s) at hand, this method facilitates the exploration of complex meanings and relationships within the data, going beyond surface-level observations to capture deeper patterns [3-5, 7]. Reflexivity and an iterative approach are essential throughout this process to ensure the validity of thematic findings [7].

Thematic analysis is crucial to informing treatment development and monitoring processes in clinical trials and mental health care [3, 4, 7, 8]. However, the manual nature of analysis is labour-intensive and poses a major barrier to scaling qualitative research in definitive clinical trials and timely personalization and co-design of digital interventions. These challenges often lead to small sample sizes, limiting generalizability and constraining the examination of individual differences across diverse populations. This issue is particularly significant in digital mental health clinical trials and VR-based interventions, where patient experiences are crucial for patient-centric co-design and personalization, as well as treatment feasibility and effectiveness. Automating aspects of thematic analysis could help fill this gap, making qualitative insights more accessible for intervention development and evaluation [8-12].

Advancements in machine learning (ML) and natural language processing (NLP) offer promising solutions [9, 13-15]. NLP-assisted coding has been incorporated into some qualitative software (e.g., NVivo), demonstrating the potential for automatic theme detection [8]. However, most existing methods rely on supervised ML approaches, which require predefined training data and are often ineffective for small, unstructured datasets commonly found in qualitative research [16-18]. While these models show promise for large-scale thematic classification, they struggle with nuanced interpretation adaptability across different research contexts [19]. LLMs represent a transformative innovation, offering more advanced contextual understanding and generative capabilities than traditional NLP techniques relying on internet-scale datasets [20, 21]. Unlike rule-based or keyword-matching approaches, such as n-gram [22] and long short-term memory [23] that operate on limited context windows or require handcrafted feature engineering, LLMs use transformer architectures to analyze large conversational contexts (e.g., 125k tokens for GPT-4o), recognizing subtle linguistic patterns and emergent themes [14]. This enables more comprehensive thematic identification, reducing manual coding efforts, and improving the analysis of complex, unstructured data [14, 24, 25]. LLMs can facilitate code generation and operationalization, excerpt extraction, and theme synthesis [9, 13, 18, 24, 25], making them particularly well-suited to qualitative research [15, 26-28], as well as psychiatry and healthcare applications [19, 29-33], potentially redefining how qualitative research is conducted in mental health clinical trials.

Despite their potential, LLM integration into qualitative research remains underexplored, due in combination to methodological, ethical, and technical barriers. Methodologically, concerns exist regarding the transparency and reproducibility of AI-assisted coding, as LLMs function as "black boxes" making it difficult to trace decision-making processes [33, 34]. Comparative studies have demonstrated that while GPT-4 can generate structured thematic outputs, it often lacks the nuanced contextual understanding achieved by human researchers [19]. Ethically, issues related to bias in model outputs, data privacy, and the misinterpretation of sensitive topics require careful oversight [31]. Technical training, high computational costs, the need for domain-specific fine-tuning, and the risk of overfitting to training data pose additional challenges [19]. Furthermore, while NLP methods have been integrated into qualitative research tools for education and the social sciences [18], there is a notable gap in empirical studies validating the reliability and applicability of LLMs for smaller, semi-structured interview datasets in mental health care.

Recent research highlights the potential role of LLMs in thematic analysis of qualitative healthcare data but underscores significant limitations. For instance, patient and participant perspectives have been evaluated using LLMs, demonstrating that they can efficiently process large interview datasets and provide thematic classification comparable to human performance [29, 33]. However, while LLM-generated themes appear to be structurally sound, they lack deep contextual understanding, requiring human oversight and validation to correct biases and errors [27, 35, 36]. GPT-4o is capable of generating coherent themes in healthcare interview data but sometimes miss nuanced patterns identified by human researchers [19]. Furthermore, while AI-assisted qualitative research is gaining traction in primary health care and psychiatry [29, 32], LLMs are not yet widely adopted.

Using interviews from a stress-reduction intervention clinical trial in healthcare workers [37, 38] as a case example, this paper introduces an innovative comparative study applying GPT-4o LLM-based thematic analysis with validation against traditional human methods. To our knowledge, this is the first study to benchmark out-of-the-box and knowledge-base LLM-generated thematic analysis against human-led thematic analysis in a digital mental health clinical trial setting, providing empirical evidence on the feasibility of integrating LLMs into qualitative digital health research and important insights into coding saturation. Using both out-of-the-box and knowledge-based LLM variations ensures a comprehensive evaluation of its capabilities: the former provides insights into the model's raw reasoning abilities, while the latter leverages domain-specific knowledge to enhance contextual accuracy [15]. The findings will contribute to the ongoing discourse on AI-human collaboration, offering critical insights into how LLMs can enhance thematic analysis while maintaining methodological rigour.

Automating qualitative coding has the potential to accelerate research and medical advancements in digital mental health and VR-based interventions, enabling faster analysis of patient experiences, informing intervention refinements, and ultimately shaping future clinical trial methodologies. Furthermore, the insights gained from this study will help establish best practices for AI-assisted qualitative research in digital healthcare and psychiatry, contributing to the responsible and ethical integration of LLMs into medical research and practice.

## II. METHODS

### A. Study Design

This comparative study evaluates the effectiveness of two GPT-4o approaches (out-of-the-box and knowledge-base) for conducting thematic analysis relevant to digital mental health, benchmarked against human-centric analysis (Fig 1).

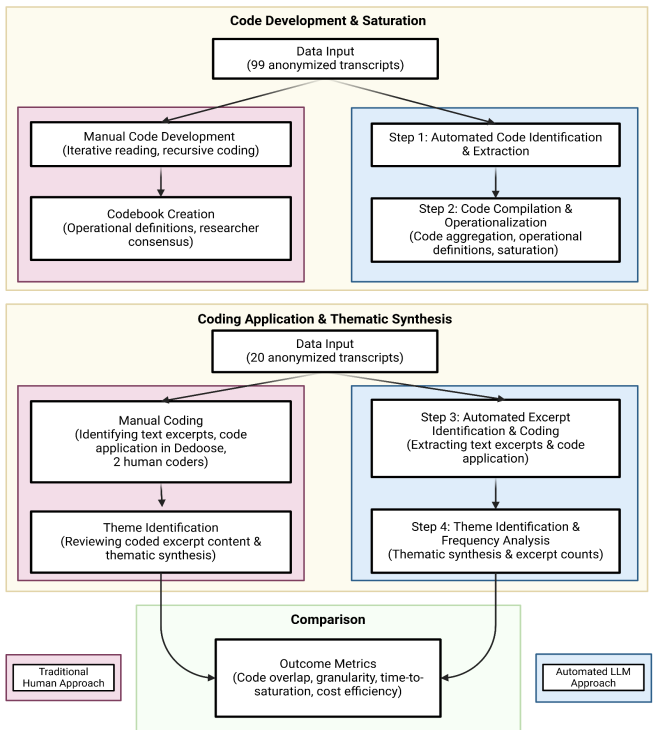

Fig. 1. Methodological process for the proof-of-concept comparison between traditional (human-based; red left-hand panel) and LLM approaches (out-of-the-box and knowledge-base; blue right-hand panel) to thematic analysis. Created in BioRender. Bhat, V. (2025) https://BioRender.com/3zeeh2d

### B. Data Source

Semi-structured interview transcripts from the VR debrief component of the DHMI-S trial in healthcare workers [37, 38] were used as a case example. Ethical approval for the DHMI-S trial (NCT05923398) was obtained in April 2023. In compliance with data privacy regulations, interviews were transcribed verbatim and anonymized using a multi-step AI-assisted transcription process (Fig. 2) to protect participant confidentiality. LLM input data did not contain any personally identifiable information.

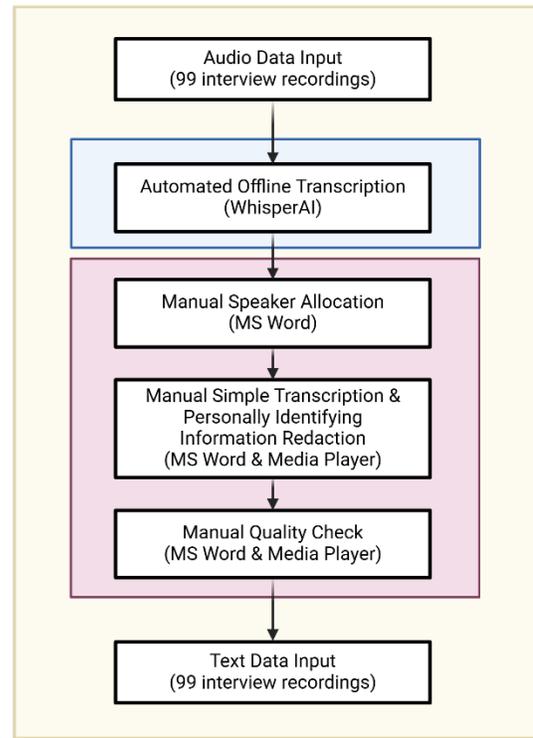

Fig. 2. AI-assisted transcription process for VR debrief interviews from the DHMI-S trial [37, 38]. WhisperAI speech-detection LLM algorithms automatically extracted text from audio files, followed by manual speaker allocation, transcription review, and quality check by research assistants trained in simple transcription [39] and anonymization conventions. Created in BioRender. Bhat, V. (2025) https://BioRender.com/srhvxog

### C. Use-Case Research Questions

VR debrief interviews from the DHMI-S trial [37, 38] were selected as use-case examples for this study because they captured detailed participant experiences, reflections, and feedback following involvement in a stress-related VR simulation and psychoeducational digital intervention, with content and research questions relevant and generalizable to VR applications and digital mental health research more broadly.

The five research questions derived from the DHMI-S interviews used as proof-of-concept case examples were:

*RQ1:* What were the sources/causes and targets of the emotions experienced in VR?

*RQ2:* When was (the most) stress experienced? How did this impact participants' decisions and behaviours?

*RQ3:* What did participants find most/least realistic about the VR simulation?

*RQ4:* What feedback do participants have to improve future VR simulations of moral distress in healthcare workers?

*RQ5:* What factors, experiences, beliefs, and/or motivations drove patient selection and clinical decision-making in the care of the avatar patients?

Direct comparison of the thematic analysis outputs produced by human researchers, an out-of-the-box LLM, and a knowledge-base LLM will provide empirical evidence informing the reliability and suitability of applying LLM-based thematic analysis in digital mental healthcare clinical trials. This paper focuses on comparative outcome measures across techniques (see below); qualitative findings will be reported elsewhere.

### D. Human-Based Thematic Analysis

Reflexive thematic analysis was conducted by a qualitative research expert, in line with Braun and Clarke's methodology [7], see Fig. 1.

First, deductive parent codes and inductive child codes were developed based on the five aforementioned research questions. All (N = 99) transcripts were reviewed, identifying recurring concepts and ideas emerging from participant responses, and updating code structure and content as appropriate. Code development continued until thematic saturation was reached (i.e., no new parent or child codes emerged from the transcripts), or all transcripts were reviewed, whichever occurred first. The number of transcripts required to achieve the finalized coding structure and operational definitions indexed the saturation point. Each parent and child code was operationally defined in a comprehensive coding manual. Coding development input was provided by another qualitative researcher prior to commencing excerpt extraction and code application.

Next, a random subset of 20 VR debrief interview transcripts from the DHMI-S trial was selected. These transcripts were manually coded in Dedoose [40] using the developed coding schema by two human researchers. Each relevant statement or excerpt within the transcripts was tagged with the appropriate code(s); coding was completed independently by two human researchers familiar with the coding schema (Cohen's κ = 0.84).

Finally, content within each code was reviewed to identify overarching and recurring themes. These themes were synthesized to provide insights into participants' perspectives related to the research questions.

### E. LLM-Based Thematic Analysis

Two LLM-based approaches were applied using OpenAI's GPT-4o: one out-of-the-box model and one knowledge-based model aligned with Braun and Clarke's [7] framework. Both used the RISEN prompt engineering strategy (Role, Instructions, Steps, End Goal, Narrowing) [41]; the RISEN framework is outlined in Fig 3 and prompts used for each step of the LLM thematic analysis are in Supplementary 1.

**Fig. 3.** Description of the RISEN prompt engineering framework. Created in BioRender. Bhat, V. (2025) https://BioRender.com/pd1cnzr

RISEN offers a structured yet flexible approach tailored to qualitative research, enabling targeted, goal-driven outputs. Compared to alternative strategies—few-shot prompting [42], which lacks adaptability; Chain-of-Thought [43], which emphasizes stepwise reasoning without outcome alignment; and Tree-of-Thoughts [44], which supports exploration at the expense of conciseness—RISEN balances relevance and precision.

LLM-based thematic analysis was conducted using GPT-4o, am OpenAI transformer-based model that captures contextual relationships through multi-head self-attention and feed-forward layers. This architecture supports a nuanced understanding of long-form text, making it suitable for qualitative analysis. The knowledge-base approach further integrated Braun and Clarke's [7] framework using Retrieval-Augmented Generation (RAG) [45], enhancing alignment with qualitative mental health research standards. This hybrid method leveraged GPT-4o's contextual capabilities while reinforcing methodological rigour in theme development.

#### 1) Model Design

The out-of-the-box GPT-4o model (accessed in August 2024) was used without the integration of formal qualitative frameworks. Its transformer-based architecture enables contextual reasoning and thematic inference across unstructured text, surpassing rule-based or traditional machine-learning NLP systems to analyze context holistically across entire datasets. This integration allows GPT-4o not only to identify linguistic structures and categorize thematic codes but also to infer deeper contextual relationships and recurring themes within unstructured text, setting it apart from traditional rule-based or machine-learning NLP systems. In contrast, the knowledge-based model incorporated Braun and Clarke's [7] methodology using RAG [45], which allows the model to retrieve and integrate relevant qualitative principles without altering underlying weights. Unlike fine-tuning—which adjusts model parameters but can compromise safety alignment—RAG maintains flexibility, ensures safety alignment, and reduces resource intensity, making it well-suited for qualitative research.

#### 2) Four-Step Thematic Analysis Process

Out-of-the-box and knowledge-base approaches both followed a structured four-step zero-shot analysis mirroring traditional thematic analysis (Fig 1). Coding development was accomplished through the first two steps: parent and child codes were generated based on individual reviews of all (N = 99) transcripts (Step 1) and compiled in batches of five transcripts to identify recurring codes and saturation point (Step 2). Thereafter, Step 3 automatically extracted excerpts and applied codes based on the coding scheme derived in Step 2 and overarching themes were identified based on excerpt synthesis (Step 4). The same prompts were used for both LLMs, all that changed was the inclusion/exclusion of the knowledge base.

#### 3) Computational Resources

The application programming interfaces (APIs) provided by OpenAI were used to access the GPT-4o model allowing us to leverage their powerful infrastructure. These infrastructures were accessed using API calls from a local machine that is a 3GHz Dual-Core Intel Core i7 processor with 16 GB 1600 MHz DDR3 memory and an Intel Iris 1536 MB graphics card. Python programming language was used for the API calls.

### F. Outcome Variables

To assess each method's effectiveness, we examined four domains: resource use, thematic saturation, output comparison, and analytic quality. Output comparisons were conducted for each stage of the process (code development, excerpt extraction and coding, and thematic synthesis). Resource use included software, personnel, and cost. Saturation was tracked by the number of transcripts needed to reach coding stability. Code, excerpt, and theme comparisons evaluated overlap, consistency, and differences in operational definitions. Reliability and validity were assessed by benchmarking LLM outputs to human coding, focusing on accuracy, consistency, and potential omissions. Thematic analysis results will be reported separately.

## III. RESULTS

### A. Resources

A full cost breakdown and comparison is provided in Supplementary 2.

The human-led thematic analysis required approximately 110 hours to complete. This included 92 hours for code development and operationalization—comprising a full review of all 99 transcripts and collaborative discussion among qualitative experts. An additional 10 hours were spent by coders identifying excerpts and applying codes in Dedoose [40], with each transcript taking an average of 30 minutes. Theme synthesis and excerpt review took another 10 hours. The total personnel cost for human-based analysis was $3,537 CAD, with Dedoose licensing adding $74.24 CAD. In comparison, the LLM-based analyses were completed by a single postdoctoral researcher over approximately 40 hours. This included prompt design, model execution, and output organization, corresponding to a personnel cost of approximately $1,260 CAD. The OpenAI API usage incurred a technical cost of $12.10 CAD.

### B. Code Development

Recursive code development by a trained qualitative researcher resulted in 7 parent codes and 65 child codes aligned with the five use-case research questions (Fig. 4 & 5A). Most research questions were linked to 1–2 parent codes, each with 6–11 child codes (M = 8.13). One parent-child code structure (Impact of Stress on Nurse Behaviour) spanned two research questions, with the parent code and 6 of its 7 child codes addressing RQ2 and one child code further informing RQ5 (Fig. 4). Saturation was reached early (within 20–30 participants) for deductive codes and frequently occurring responses, particularly when parent codes had fewer subcategories (e.g., emotional sources or targets). For more distributed codes—such as those related to VR realism—saturation required reviewing all (n = 99) transcripts.

The out-of-the-box model produced a set of parent and child codes addressing all research questions (Fig 4). There were no major differences in codes derived from the out-of-the-box and knowledge-base approaches beyond minor differences in code phrasing (e.g., Emotional Sources vs. Sources of Emotion). In total, 10 parent codes and 22 child codes were identified (Fig. 5A), with 2 or 3 child codes per parent code.

Early codes (e.g., Emotions, Improved Suggestions) emerged within the first 5–10 transcripts and became more refined as additional data (in increments of five participants) was reviewed by each LLM. The out-of-the-box model reached thematic saturation between 15–20 participants, with no new codes appearing beyond 15. The knowledge-based model achieved saturation more quickly, between 10–15 participants. After this point, parent and child codes remained consistent, with only minor wording adjustments (e.g., Emotional Sources vs. Sources of Emotion).

Of the 10 parent codes generated by the LLMs, four aligned directly with human-defined codes: Emotional Sources, Emotional Targets, Stress Timing, and Behavioural Impact of Stress. Four VR-related parent codes from the LLMs (each with two child codes) reflected different aspects of simulation realism, whereas the human coder grouped these under a single parent code with 11 child codes, suggesting a more integrated thematic structure. The LLMs also introduced a parent code related to participants' beliefs and ethical decision-making, which was conceptually present but organized differently in the human coding structure. All LLM-generated codes were conceptually represented in the human codes developed in consideration of the five use-case research questions, with the exception of *Ethical Considerations* (Fig 4). Alternatively, the human-derived *Patient Selection* code was only loosely reflected by the LLM's Patient Condition child code, indicating partial overlap rather than direct correspondence.

### C. Excerpt Identification and Coding Application

Across 20 participants, 428 excerpts were identified by human researchers (range: 14–37 per participant; M = 21.4; Fig. 5B), with 190 (44%) not captured by either LLM. Human-derived parent codes contained between 1–19 excerpts (M = 8.5), while child codes ranged from 0–36 (M = 9.19). Manually-selected excerpts varied in length (3–364 words; M = 49.69) and often included contextual features such as interviewer clarifications, interjections, and disfluencies (e.g., "uhm," stuttering).

The out-of-the-box model extracted 417 excerpts across 20 participants (range: 14–35; M = 21; Fig. 5B), with 88 (21%) uniquely identified by the model. In contrast, the knowledge-based model only extracted 251 excerpts (range: 8–25; M = 13), with 35 (14%) uniquely identified. In both cases, the excerpts were exclusively assigned to child codes (out-of-the-box LLM: range = 5–41 per code; M = 19; knowledge-base LLM: range = 2–26 per code; M = 11); none were classified at the parent-code level. LLM-generated excerpts were brief (out-of-the-box LLM: range = 3–78 words; M = 22; knowledge-base LLM: range = 5-63, M =18) and included minimal interviewer context.

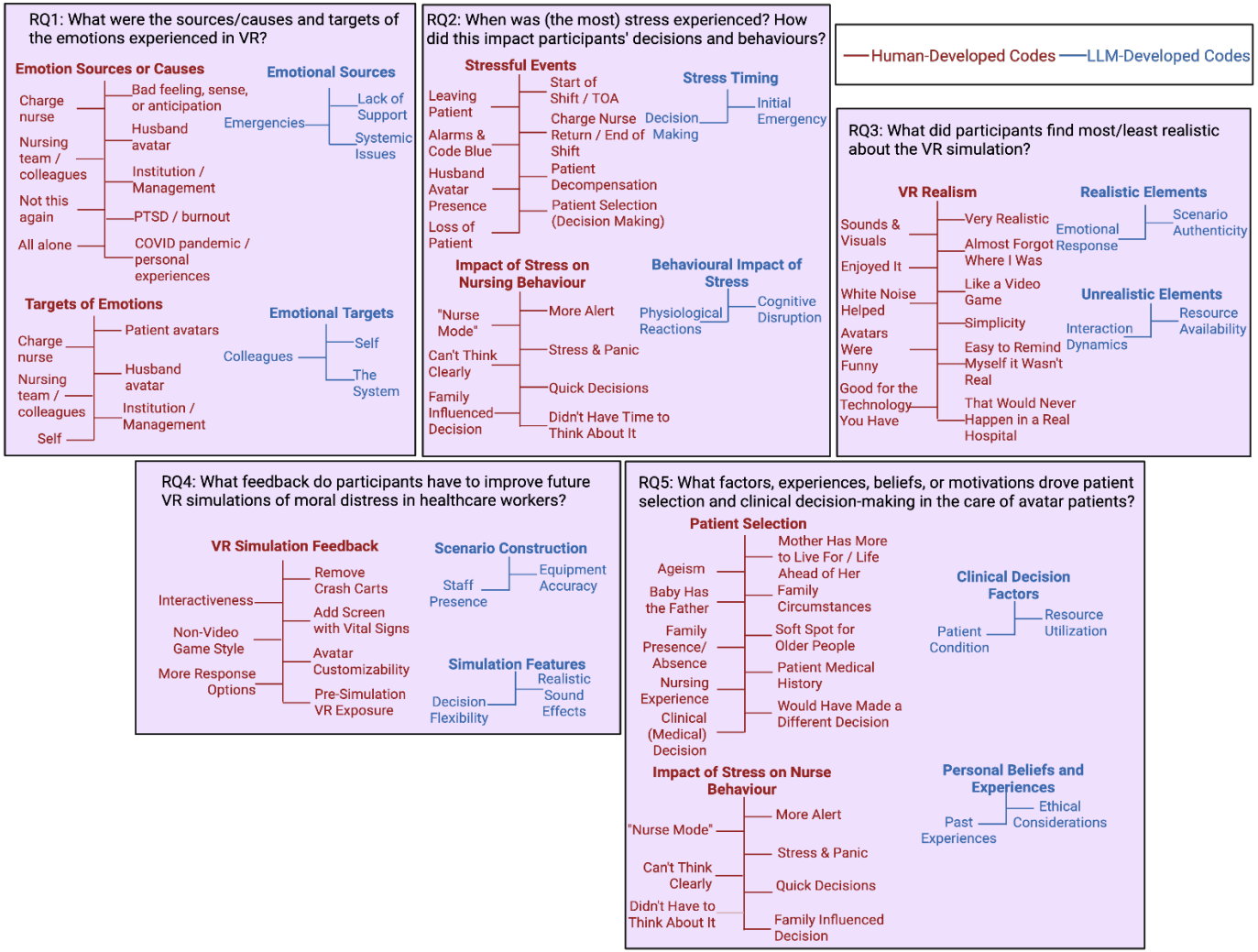

Fig 4. Distribution of code outputs for human (red) and LLM (blue) approaches across dataset-specific research questions. Created in BioRender. Bhat, V. (2025) https://BioRender.com/ghyv39m

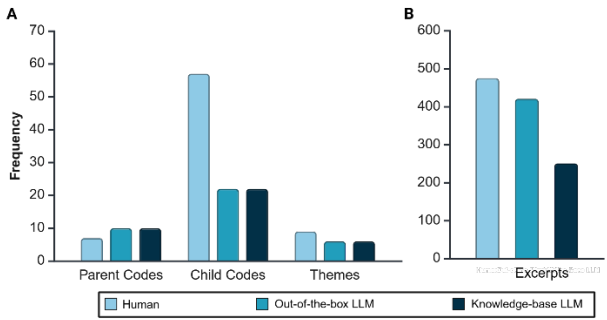

Fig 5. Histograms depicting frequency counts of human, out-of-the-box GPT-4o LLM, and knowledge-based GPT-4o LLMs. Panel A: Frequency comparison of codes and themes. Panel B: Frequency comparison of excerpts. Created in BioRender. Bhat, V. (2025) https://BioRender.com/tolmebo

Overall, 238 human excerpts (56%) were also identified by the out-of-the-box LLM, while the knowledge-based LLM matched 124 human excerpts (29%). Out-of-the-box excerpts largely overlapped with human ones: 73% were fully embedded within longer, more context-rich human excerpts (81%). A smaller portion of human excerpts (14%) were shorter than LLM counterparts. Twenty-four (6%) of the out-of-the-box excerpts were only partially captured by humans, and 12 (5%) were exact matches. Minor output errors included 11 excerpts (2.6%) containing only interviewer dialogue, 2 (0.48%) irrelevant responses, 3 (0.72%) that replaced disfluencies with ellipses, and 5 (1.2%) that removed filler words without indication. The knowledge-based LLM matched human-selected excerpts in 54% of cases, with 4% only partially overlapping. However, 68 excerpts (27%) were interpretive summaries rather than direct quotes, and 2 excerpts (1%) contained only interviewer speech.

In all three cases, the majority of excerpts were single coded (human: 62%; out-of-the-box LLM: 78%, knowledge-base LLM: 91%), with fewer excerpts being dual-coded (human: 26%; out-of-the-box LLM: 18%, knowledge-base LLM: 5%) and a minority having three or more codes applied (human: 12%; out-of-the-box LLM: 5%, knowledge-base LLM: 3%).

### D. Theme Synthesis

Human researchers identified nine overarching themes; the out-of-the-box and knowledge-base LLMs each procured six themes (Fig. 5A & 6).

Substantial thematic overlap was observed between LLM approaches (Fig. 6). All out-of-the-box themes corresponded with at least one theme from the knowledge-based LLM, with four showing direct 1:1 matches and two aligning with two knowledge-based themes. In contrast, one-third of human-identified themes were absent from both LLM outputs, while another third mapped to two out-of-the-box themes but only one from the knowledge-based model. The remaining themes either aligned with both models (22%) or matched two knowledge-based themes but only one out-of-the-box theme (11%).

Human analysis prioritized emotional distress and personal context, while LLMs focused more on systemic and procedural dimensions of decision-making and ethical considerations. Human coders also provided detailed feedback on VR realism and its relevance to COVID-19, whereas LLMs offered broader reflections on stress and training. Nonetheless, both human and LLM approaches consistently identified key themes such as the impact of systemic factors, emotional burden, ethical reasoning, and the utility of VR training, along with shared participant recommendations for improving realism and alignment with real-world scenarios.

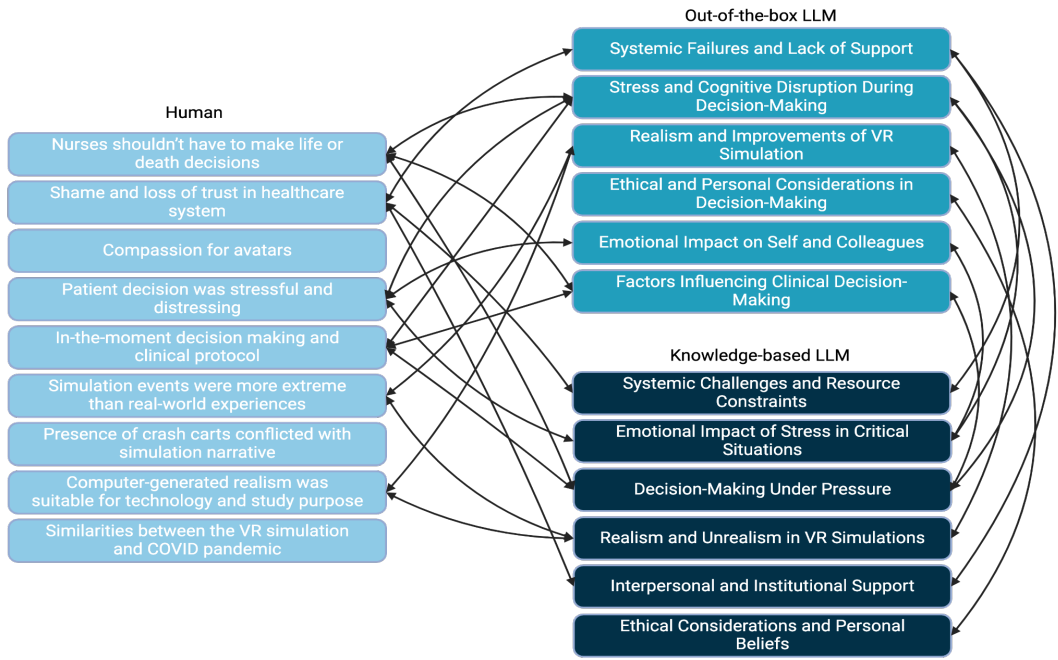

**Fig 6.** Themes generated by human researchers, out-of-the-box GPT-4o LLM, and a knowledge-base GPT-4o LLM with conceptual overlap indicated with bidirectional arrows. Created in BioRender. Bhat, V. (2025) https://BioRender.com/dzfaore

## IV. DISCUSSION

This study compared the effectiveness of out-of-the-box and knowledge-base GPT-4o LLM thematic analysis processes in code development, excerpt extraction, and thematic synthesis benchmarked against traditional human-based analysis in the context of digital healthcare research. By analyzing transcripts from 20 participants using three different approaches we sought to explore the consistency, depth, and practicality of these methods in generating meaningful insights, particularly regarding emotional experiences, stress factors, perceived realism of the VR simulation and decision-making.

The out-of-the-box LLM generated broad parent and child codes, offering a general overview of participants' responses. Although saturation was achieved quickly (within 20 transcripts), codes often lacked depth and nuance. The knowledge base-guided LLM produced similar codes as the out-of-the-box model but was able to achieve saturation with fewer participants, suggesting improved efficiency. While both LLMs demonstrated moderate consistency with human-derived codes, they struggled to capture the depth critical for capturing individual differences and contextual nuances, consistent with recent reports in marketing [34], education [27], and healthcare [33]. Traditional human-based code development excelled in identifying nuanced emotions and personal experiences, offering richer interpretations but requiring significantly more time and resources.

Notably, LLM approaches differed considerably from human researchers in excerpt extraction and coding application. While the out-of-the-box LLM extracted a comparable number of excerpts to humans, 21% of automatically-extracted excerpts were unique to the out-of-the-box LLM model and 44% of manually-extracted excerpts were not captured by either LLM. The knowledge-based LLM was markedly inferior in the excerpt stage of analysis, underscoring concerns about accuracy and representativeness in LLM-generated excerpts. Human-selected excerpts were also notably longer and richer in contextual and emotional cues, often including interviewer interjections and disfluencies that were largely excluded in LLM outputs. The excerpt coding comparison further confirmed this difference in granularity. While human excerpts were frequently multi-coded, LLM excerpts were primarily coded with a single theme, reinforcing the more reductive nature of LLM-based interpretations.

Despite these differences, all methods identified overlapping high-level themes relevant to digital mental health research, including emotional distress, decision-making complexity, and the perceived realism of VR simulations. LLM-generated themes were more systemically focused and generalized, while the human analysis emphasized specific emotional experiences with more precise links to contextual factors such as the COVID-19 pandemic. human researchers identified three additional themes not captured by either LLM approach and provided more operationalized distinctions between overlapping concepts than LLMs. It is notable that the LLMs captured the influence of ethics and morals in child code development and theme synthesis when prompted with five use-case research questions, whereas human codes (developed in consideration of RQ 1-5; Fig. 4) do not reflect these attributes. However, within the larger thematic analysis of the rich DHMI-S trial interview data [37] – to be reported elsewhere – the parent code *Violating (or not) Morals, Values, and Beliefs* was developed in consideration of a sixth research question: *Did the VR simulation violate participants' morals, values, or beliefs? If so, how?* This divergence illustrates that while LLMs can reliably identify broad patterns and recurring themes, they are less attuned to the fine-grained emotional and contextual elements that often drive clinical qualitative insights [18, 27, 32-34].

### A. Key Findings

Alignment across codes, excerpts, and overarching themes generated by human and LLM approaches underscores the potential of LLMs to reliably identify key patterns in semi-structured interviews relevant to digital healthcare trials. Consistent with emerging research [18, 24, 36], LLMs provided broad overviews of big-picture codes and themes, whereas the human-based thematic analysis uncovered greater specificity, such as detailed emotional triggers linked to avatars or events in the VR simulation. These intricacies were less consistently captured by LLM-based methods, highlighting their limitations in matching the depth of human-driven analysis [33, 34, 36].

The knowledge-base guided LLM approach stood out for its efficiency in code development, achieving thematic saturation with fewer participants but significantly underperformed on excerpt identification and code application. The out-of-the-box LLM was most consistent with human researchers across all stages of thematic analysis, although some limitations persisted. The GPT-4o structure enhanced the LLMs' abilities to generate contextually grounded and detailed codes, excerpts, and themes relevant to digital mental health, demonstrating the potential of LLMs as a cost-effective method for aiding qualitative researchers when resources are limited. Consistently achieving code saturation in fewer than 20 transcripts further demonstrates LLM suitability for smaller-scale interview datasets, such as those incorporated in pilot trials where participant feedback is crucial to patient-centric treatment.

Overall, these findings emphasize the utility of LLM-based methods in thematic analysis while highlighting areas where traditional human-based approaches maintain an advantage, particularly in capturing the depth and nuance of participants' lived experiences. The structured application of thematic frameworks to LLMs represents a significant step forward in AI advancement of healthcare analytics, suggesting opportunities for further refinement and integration of these methods in future research.

### B. Implications for Healthcare Research

LLMs present a transformative and timely opportunity to revolutionize mental healthcare research by automating core components of thematic analysis, including code development, excerpt extraction, and theme synthesis. These tools enable more scalable, rapid, and cost-effective approaches to qualitative analysis, which is particularly valuable for large-scale studies where manual coding is labor-intensive, as well as smaller-scale pilot studies with limited transcripts [2, 3, 15, 26, 30, 32, 33]. The urgency for such scalable solutions is heightened by increasing mental health needs globally, where timely insight generation is essential to inform responsive interventions, clinical workflows, and public health policy.

LLMs may also help revolutionize decision support tools and patient-facing technologies by enabling more adaptive, emotionally intelligent systems capable of parsing large volumes of narrative data and translating it into actionable insights [29, 32, 33]. However, these benefits are not without caveats [30-32]. Although LLMs—particularly out-of-the-box models—perform well in generating broad codes and over-arching themes, they tend to overlook critical contextual nuances, emotional inflections, and multi-layered meaning that are essential in mental health settings [32] and under-perform at excerpt identification and code application relative to human researchers. By embedding LLMs into structured thematic frameworks and pairing them with expert human review, researchers can optimize both efficiency and analytical rigor – see [34] for applications outside healthcare. This hybrid model holds significant promise for generating rich, timely, and actionable insights that can guide clinician decision-making, inform patient-centered care, and accelerate innovation in mental healthcare delivery systems [30, 33].

### C. Limitations

This study evaluated LLM-assisted thematic analysis using GPT-4o in both out-of-the-box and knowledge-base guided forms. Findings may not generalize to other models (e.g., DeepSeek, LLaMa, Gemini) or alternative prompting techniques such as chain-of-thought or tree-of-thought reasoning. Differences in model architecture, training data, and prompt design can significantly influence both output quality and interpretability. Furthermore, despite highlighting the benefits of AI-assisted thematic analysis for mental health care, the study's focus on healthcare professionals' limits generalizability to patient populations. Applying LLMs to patient interview data—particularly in mental health research—raises heightened ethical and privacy concerns due to the sensitivity of the content, the vulnerability of participants, and the potential for inadvertent disclosure of identifiable or stigmatizing information. Unlike interviews with professionals, patient narratives often involve deeply personal accounts that require strict adherence to confidentiality standards and data protection protocols. Therefore, future work should be guided by clearly defined ethical oversight and governance frameworks when incorporating LLMs into analyses involving patient data.

LLM performance in generating meaningful codes and themes is further influenced by training data and prompt engineering (13-15, 33). As LLMs evolve, their capabilities are likely to improve, but researchers must remain mindful of these dependencies (33) and their environmental impact, particularly the significant carbon footprint associated with training and inference. While the LLMs produced relevant high-level codes and was more cost-effective, it lacked the depth of human analysis with non-trivial environmental costs for LLM deployment. Another limitation is the potential for bias in LLM-generated themes, stemming from biases present in their training datasets. These biases may result in the omission or skewing of certain perspectives, particularly when datasets lack socio-cultural diversity. For instance, LLMs trained on culturally narrow datasets may overlook or misrepresent marginalized perspectives. Output consistency can also vary between models, complicating reproducibility.

To mitigate such risks, researchers should employ cross-validation methods on diverse datasets, and benchmark LLM outputs against human-coded themes. These practices can enhance the credibility, inclusivity, and ethical integrity of LLM-driven thematic analysis in both professional and patient-centred research.

### D. Future Directions

This study highlights several promising avenues for advancing AI-assisted qualitative research, primarily developing and evaluating hybrid approaches that integrate LLMs with traditional human-led thematic analysis. Such models could capitalize on the efficiency and scalability of LLMs while ensuring the interpretive depth, contextual awareness, and methodological rigour provided by human analysts in mental health care. Further research should also assess LLM performance across diverse patient populations and mental health conditions beyond stress. While this study focused on healthcare professionals, applying LLM-assisted analysis to patient interviews—particularly in areas such as depression, anxiety, trauma, substance use, and chronic illness—could uncover new insights into lived experiences, treatment engagement, and recovery trajectories. These contexts involve complex, emotionally nuanced narratives that challenge automated analysis, making them critical for evaluating the sensitivity and clinical relevance of LLM outputs.

Likewise, systematic evaluation of prompt engineering strategies will be necessary to inform which aspects of the RISEN framework truly enhance thematic accuracy and contextual sensitivity. Ablation studies on the RISEN framework may help clarify which steps are essential for optimal performance across different types of qualitative data. Contextualization of the RISEN approach relative to other frameworks and benchmarking across LLMs (e.g., GPT-4o, Claude, DeepSeek, LLaMa, Gemini) is also warranted. Differences in architecture and training could influence each model's ability to generate nuanced, methodologically coherent codes and themes. Comparatively evaluating these models in terms of thematic saturation, consistency, and fidelity to source data will inform future model selection for qualitative research.

Finally, there is significant potential in developing framework-aligned LLMs tailored to established qualitative methodologies, such as Braun and Clarke's thematic analysis [7]. Fine-tuning models to adhere to such frameworks could support more rigorous, context-sensitive, and interpretable AI-assisted analysis—paving the way for meaningful integration of LLMs in qualitative research workflows.

## V. Conclusions

Overall, these findings emphasize both the promise and the current limitations of LLM-assisted thematic analysis. The knowledge base-guided LLM offered improvements in code development, saturation efficiency, and contextual alignment relative to the out-of-the-box model, but still underperformed in comparison to human analysis in terms of the depth, specificity, and accuracy. A hybrid model—where LLMs support initial code development and preliminary theme construction, with human oversight for excerpt identification, coding application, and theme refinement—may offer the most balanced approach, consistent with recent applications in marketing research [34]. This would leverage LLM efficiency while preserving the depth and contextual richness essential to robust qualitative analysis, harnessing the strengths of both approaches, leading to richer, more reliable findings in qualitative mental health research.

## Acknowledgment

The authors thank Amit Bhat, Annabelle Persaud, Samyah Sijal, and Melody Tjong for their contributions.